\documentclass[runningheads]{llncs}
\usepackage{amsmath,graphicx,url,times,amsmath,amssymb,acronym,graphicx,balance}
\usepackage{color,booktabs}
\usepackage[table]{xcolor}
\usepackage{cite}
\usepackage{lipsum}

\acrodef{STFT}{short-time Fourier transform}
\acrodef{MSE}{mean-squared error}
\acrodef{PSD}{power spectral density}
\acrodef{RTF}{relative transfer function}
\acrodef{SNR}{signal-to-noise ratio}
\acrodef{PDF}{probability density function}
\acrodef{DOA}{direction-of-arrival}
\acrodef{VAD}{voice activity detector}
\acrodef{MVDR}{minimum variance distortionless response}
\acrodef{AIR}{acoustic impulse response}
\acrodef{PESQ}{perceptual evaluation of speech quality}
\acrodef{STOI}{short-time objective intelligibility}
\acrodef{LSD}{log spectral distance}
\acrodef{CD}{cepstral distance}
\acrodef{WER}{word error rate}
\acrodef{SPP}{speech presence probability}
\acrodef{RNN}{recurent neural network}
\acrodef{LSTM}{long-term short-term}
\acrodef{GRU}{gated recurrent unit}
\acrodef{FF}{feed forward}
\acrodef{ReLU}{rectified linear unit}
\acrodef{GCC}{generalized cross-correlation}
\acrodef{RMSE}{root-mean-square error}
\acrodef{CPSD}{cross-power spectral density}
\acrodef{SI-SDR}{scale-invariant signal-to-distortion ratio}
\acrodef{SDR}{signal-to-distortion ratio}
\acrodef{MagMSE}{Magnitude MSE}
\acrodef{LPS}{logarithmic power spectrum}
\acrodef{CSE}{complex spectrum error}
\acrodef{MagSE}{magnitude spectrum error}
\acrodef{LogMagSE}{logarithmic magnitude spectrum error}
\acrodef{DL}{deep learning}
\acrodef{fwSegSNR}{frequency-weighted segmental SNR}

\begin{document}
\title{Data augmentation and loss normalization for deep noise suppression} 
%
%
\author{Sebastian Braun \and
Ivan Tashev}
%
\authorrunning{S. Braun et al.}
%
\institute{Microsoft Research,Redmond, WA, USA\\
	\email{\{sebastian.braun, ivantash\}@microsoft.com}\\
	\url{https://www.microsoft.com/en-us/research/group/audio-and-acoustics-research-group}}
\maketitle              
\begin{abstract}
Speech enhancement using neural networks is recently receiving large attention in research and being integrated in commercial devices and applications. In this work, we investigate data augmentation techniques for supervised deep learning-based speech enhancement. We show that not only augmenting SNR values to a broader range and a continuous distribution helps to regularize training, but also augmenting the spectral and dynamic level diversity. However, to not degrade training by level augmentation, we propose a modification to signal-based loss functions by applying sequence level normalization. We show in experiments that this normalization overcomes the degradation caused by training on sequences with imbalanced signal levels, when using a level-dependent loss function.

\end{abstract}
\section{Introduction}
Speech enhancement using neural networks has recently seen large attention and success in research \cite{Wang2018,Reddy2020} and is being implemented in commercial applications also targeting real-time communication. An exciting property of deep learning-based noise suppression is that it also reduces highly non-stationary noise and background sounds such as barking dogs, banging kitchen utensils, crying babys, construction or traffic noise, etc. This has not been possible so far using single-channel statistical model-driven speech enhancement techniques that often only reduce quasi-stationary noise \cite{Ephraim1984,Gerkmann2011,Martin2001}. Notable approaches towards real-time implementations have been proposed e.\,g.\, in \cite{Tu2018,Valin2018,Tan2018,Wichern2017,Xia2020}.

The dataset is a key part of data-driven learning approaches, especially for supervised learning. It is a challenge to build a dataset that is large enough to generalize well, but still represents the expected real-world data sufficiently. Data augmentation techniques can not only help to control the amount of data, but is also necessary to synthesize training data that represents all effects encountered in practice.

While in many publications, data corpus generation is only roughly outlined due to lack of space, or often exclude several key practical aspects, we direct this paper on showing contributions on several augmentation techniques when synthesizing a noisy and target speech corpus for speech enhancement. In particular, we show the effects of increasing the SNR range and using a continuous instead of discrete distribution, spectral augmentation by applying random spectral shaping filters to speech and noise, and finally level augmentation to increase robustness of the network against varying input signal levels.

As we found that level augmentation can decrease the performance when using signal-level dependent losses, we propose a normalization technique for the loss computation that can be generalized to any other signal-based loss. We show in experiments on the CHIME-2 challenge dataset that the augmentation techniques and loss normalization substantially improve the training procedure and the results.

In this paper, we first introduce a the general noise suppression task in Section~\ref{sec:sigmodel}. In Section~\ref{sec:network}, we describe the used real-time noise suppression system based on a recurrent network, that works on a single frame in - single frame out basis, i.\,e.\, requires no look-ahead and memory buffer, and describe the training setup. In Section~\ref{sec:loss}, we describe the used loss function and propose a normalization to remove the signal level dependency of the loss. In Section~\ref{sec:augmenation}, we describe augmentation techniques for \ac{SNR}, spectral shaping, and sequence level dynamics. The experiments are shown in Section~\ref{sec:experiments}, and Section~\ref{sec:conclusion} concludes the paper.

\section{Deep Learning Based Noise Suppression}
\label{sec:sigmodel}
In a pure noise reduction task, we assume that the observed signal is an additive mixture of the desired speech and noise. We denote the observed signal $X(k,n)$ directly in the \ac{STFT} domain, where $k$ and $n$ are the frequency and time frame indices as
\begin{equation}
X(k,n) = S(k,n) + N(k,n),
\end{equation}
where $S(k,n)$ is the speech and $N(k,n)$ is the disturbing noise signal. Note that the speech signal $S(k,n)$ can be reverberant, and we only aim at reducing additive noise.

The objective is to recover an estimate $\widehat S(k,n)$ of the speech signal by applying a filter $G(k,n)$ to the observed signal by
\begin{equation}
\label{eq:Shat}
\widehat S(k,n) = G(k,n) \, X(k,n).
\end{equation}
The filter $G(k,n)$ can be either a real-valued suppression gain, or a complex-valued filter.
While the former option (also known as \emph{mask}) only recovers the speech amplitude, a complex filter could potentially also correct the signal phase. In this work, we use a suppression gain.

\section{Network and Training}
\label{sec:network}
\begin{figure}[tb]
	\centering
	\includegraphics[width=.8\columnwidth,clip,trim=90 30 70 20]{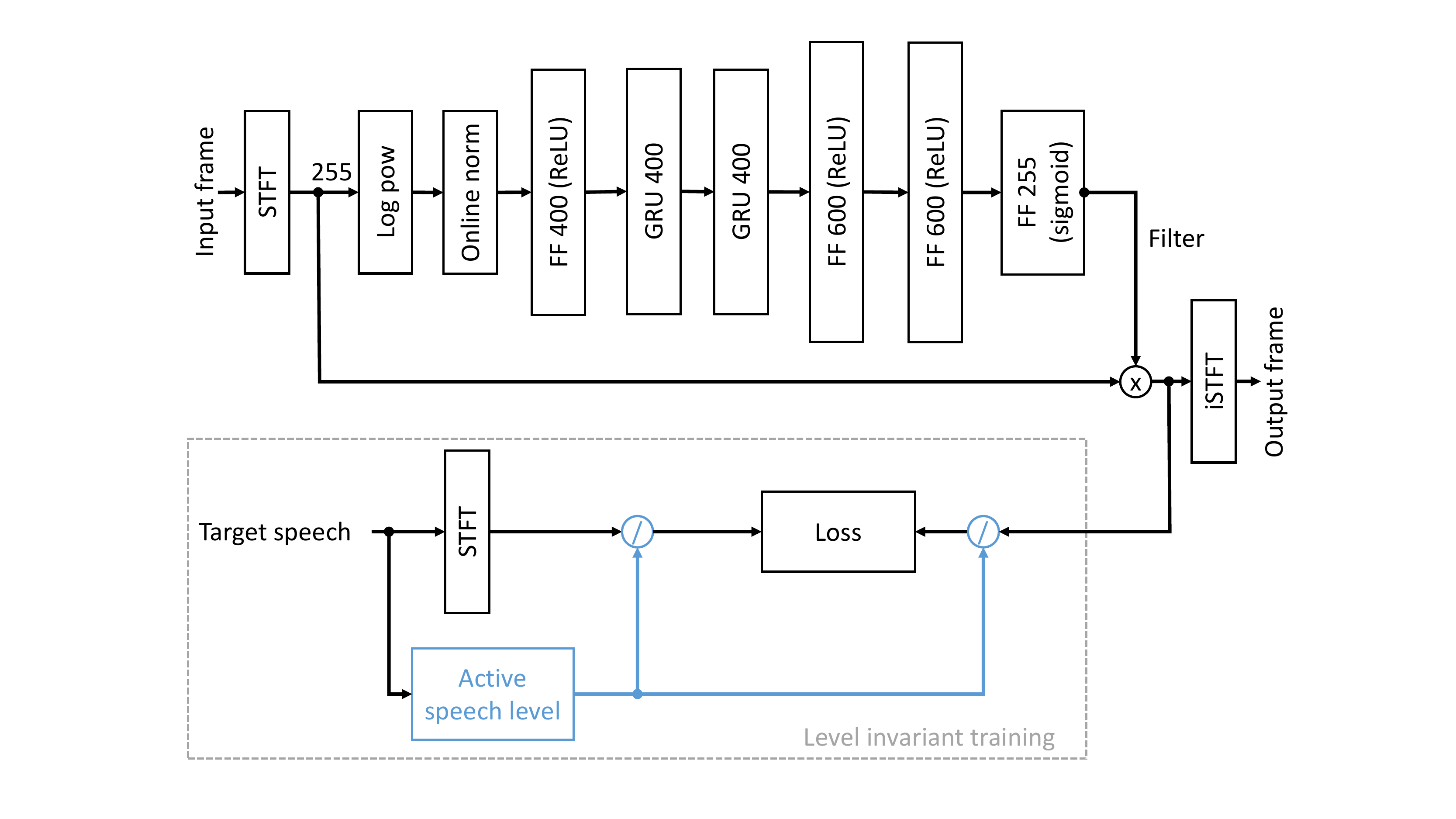}
	\caption{Network architecture and enhancement system and training procedure.}
	\label{fig:architecture}
\end{figure}

We use a rather straightforward recurrent network architecture based on \acp{GRU} \cite{Cho2014} and \ac{FF} layers, similar to the core architecture of \cite{Wisdom2019} without convolutional encoder layers. 
Input features are the logarithmic power spectrum $P = \log_{10}(|X(k,n)|^2 + \epsilon)$, normalized by the global mean and variance of the training set.
We use a \ac{STFT} size of 512 with 32~ms square-root Hann windows and 16~ms frame shift, but feed only the relevant 255 frequency bins into the network, omitting 0th and highest (Nyquist) bins, which do not carry useful information.
The network consists of a \ac{FF} embedding layer, two \acp{GRU}, and three \ac{FF} mapping layers. All \ac{FF} layers use \ac{ReLU} activations, except for the last output layer. When estimating a real-valued suppression gain, a \emph{Sigmoid} activation is used to ensure positive output. The network architecture is shown in Fig.~\ref{fig:architecture}, and has 2.8~M parameters.

The network was trained using the AdamW optimizer \cite{Loshchilov2019} with an initial learning rate of $10^{-4}$, which was dropped by a factor of 0.9 if the loss plateaued for 5 epochs.
The training was monitored every 10 epochs using a validation subset. The best model was chosen based on the highest \ac{PESQ} \cite{PESQ_P862} on the validation set.
All hyper-parameters were optimized by a grid search and choosing the best performing parameter for \ac{PESQ} on the validation set.

\section{Level Invariant Normalized Loss Function}
\label{sec:loss}
The speech enhancement loss function is typically a distance metric between the enhanced and target spectral representations. The dynamically compressed loss proposed in \cite{Ephrat2018,Wilson2018} has been shown to be very effective. A compression exponent of $0<c \leq1$ is applied to the magnitudes, while the compressed magnitudes are combined with the phase factors again. Furthermore, the magnitude only loss is blended with the complex loss with a factor $0 \leq \alpha \leq 1$.
\begin{equation}
\label{eq:loss}
\mathcal{L} = \alpha \sum_{k,n} \left||S|^c e^{j\varphi_{S}} - |\widehat{S}|^c e^{j\varphi_{\widehat{S}}} \right|^2 + (1-\alpha) \sum_{k,n} \left||S|^c - |\widehat{S}|^c\right|^2.
\end{equation}
We chose $c=0.3$ and $\alpha = 0.3$.

A common drawback of all similar signal-based loss functions is the dependency on the level of the signals $S$ and $\widehat{S}$. This might have an impact on the loss when computing the loss over a batch of several sequences, which exhibit large dynamic differences. It could be that large signals dominate the loss, creating less balanced training.

Therefore, we propose to normalize the signals $S$ and $X$ by the active signal level of each utterance, before computing the loss. The normalized loss is computed as given by \eqref{eq:loss}, but using the normalized signals $\tilde S = \frac{S}{\sigma_S}$ and $\tilde X = \frac{X}{\sigma_S}$, where $\sigma_S$ is the active speech level per utterance, i.e. the target speech signal standard deviation computed only for active speech frames. Note that this normalization does not affect the input features of the network: they still exhibit the original dynamic levels.

\section{Data Augmentation Techniques}
\label{sec:augmenation}
Especially for small and medium-scale datasets, augmentation is a powerful tool to improve the results. In the case of supervised speech enhancement training, where the actual noisy audio training data is generated synthetically by mixing speech and noise, there are some augmentation steps, which are essential to mimic effects on data encountered in the wild. Disregarding reverberation, we need to be able to deal with different \acp{SNR}, audio levels, and filtering effects that can be caused e.g. by acoustics (room, occlusion), or the recording device (microphone, electronic transfer functions).

\begin{figure}[tb]
	\centering
	\includegraphics[width=.6\columnwidth,clip,trim=290 40 250 90]{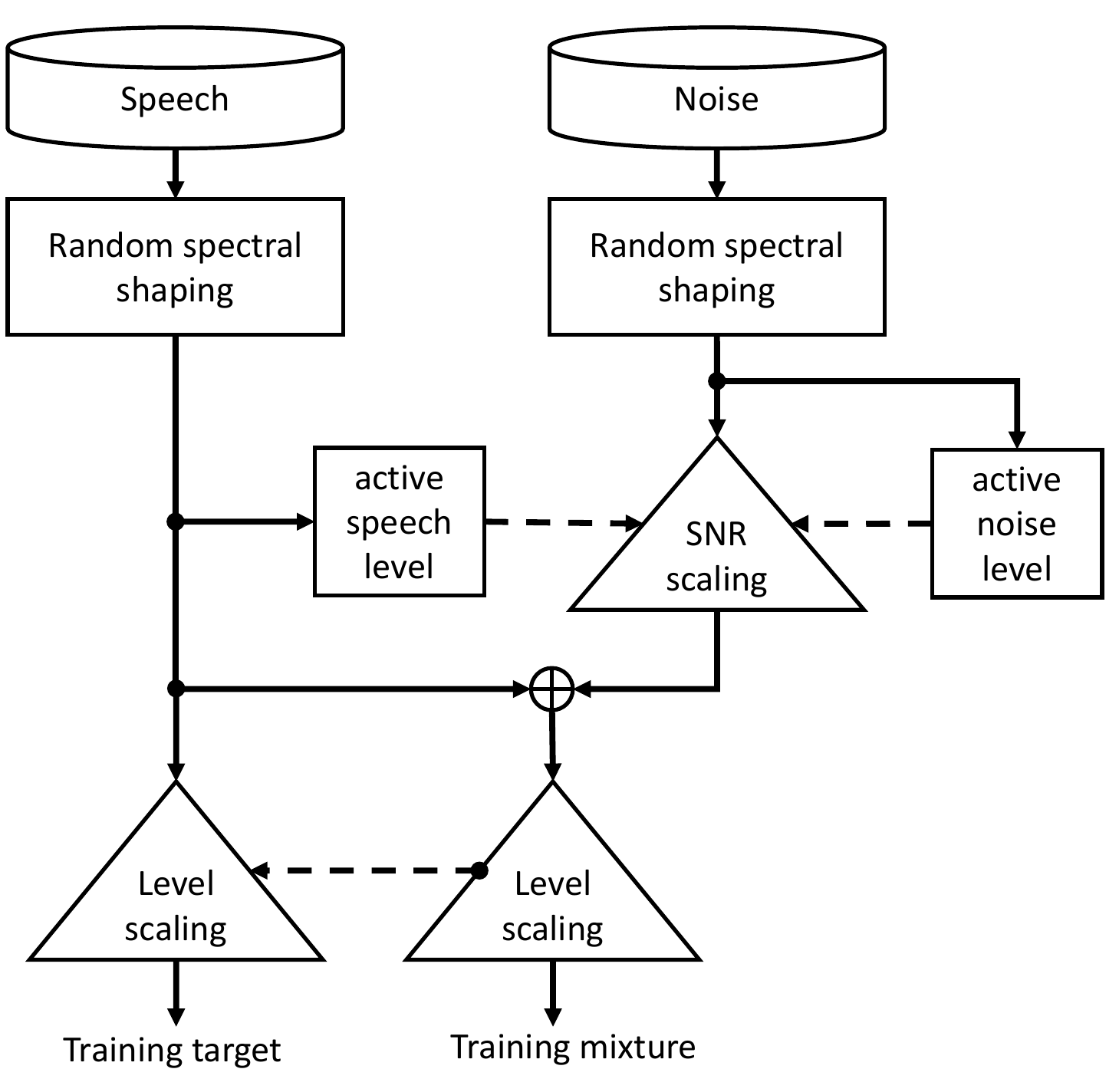}
	\caption{On-the-fly training augmented data generation.}
	\label{fig:data_generation}
\end{figure}
Our augmentation pipeline is shown in Fig.~\ref{fig:data_generation}. Before mixing speech with noise, we applied random biquad filters \cite{Valin2018} to each noise sequence and speech sequences separately to mimic different acoustic transmission effects. From these signals, active speech and noise levels are computed using a level threshold-based \ac{VAD}. Speech and noise sequences are then mixed with a given \ac{SNR} on-the-fly during training. After mixing, the mixture is scaled using a given level distribution. The clean speech target is scaled by the same factor as the mixture.
The data generation and augmentation procedure is depicted in Fig.~\ref{fig:data_generation}.

\section{Experiments}
\label{sec:experiments}

\subsection{Dataset and Experimental Setup}
We used the CHIME-2 WSJ-20k dataset \cite{Vincent2013}, which is currently, while only being of medium size, the only realistic self-contained public dataset including matching reverberant speech and noise conditions. The dataset contains 7138, 2418, and 1998 utterances for training, validation and testing, respectively. The utterances are reverberant using binaural room impulse responses, and noise from the same rooms was added with \acp{SNR} in the range of -6 to 9~dB in the validation and test sets. We used only the left channel for our single-channel experiments.

The spectral augmentation filters are designed as proposed in \cite{Valin2018} by
\begin{equation}
H(z) = \frac{1 + r_1z^{-1} + r_2 z^{-2}}{1 + r_3z^{-1} + r_4 z^{-2}}
\end{equation}
with $r_i$ being uniformly distributed in $[-\frac{3}{8},\frac{3}{8}]$.
For \ac{SNR} augmentation, the mixing \acp{SNR} were drawn from a Gaussian distribution on the logarithmic scale with mean 5~dB and standard deviation 10~dB. The signal levels for dynamic range augmentation were drawn from a Gaussian distribution with mean -28~dBFS and variance 10~dB. 

We evaluate our experiments on the development and test set from the CHIME-2 challenge. As objective metrics, we use \ac{PESQ} \cite{PESQ_P862}, \ac{STOI} \cite{Taal2011}, \ac{SI-SDR} \cite{Roux2019}, \ac{CD}, and \ac{fwSegSNR} \cite{Hu2008}.

\subsection{Results}
\begin{figure}[tb]
	\centering
	\includegraphics[width=.9\columnwidth,clip,trim=25 0 40 0]{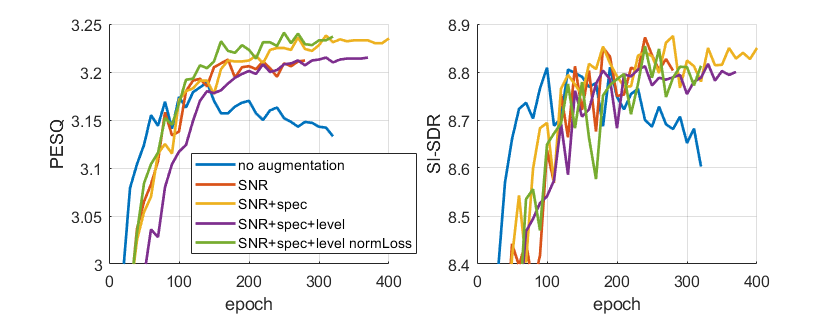}
	\caption{Validation metrics for various augmentation techniques.}
	\label{fig:dev_augmentation}
\end{figure}
Fig.~\ref{fig:dev_augmentation} shows the training progression in terms of \ac{PESQ} and \ac{SI-SDR} on the validation set. The blue curve shows training on the original data without any augmentation, mixed the 6 different \ac{SNR} levels between -6 and 9~dB. We can see that the validation metrics decrease after 150 epochs. When applying \ac{SNR} augmentation (red curve) with a broader and continuous distribution, we prevent the early validation decrease and can train for 280 epochs. Further, adding spectral augmentation increases the validation PESQ slightly. However, the level augmentation when training with standard loss \eqref{eq:loss} (purple curve) decreases the performance compared to SNR and spectral augmentation only. We attribute this effect to the large level imbalance per batch, which affects the standard level-dependent loss function. When computing the loss from normalized signals (green curve), this drawback is overcome and we obtain similar or even slightly better results than the yellow curve, but making the system robust to varying input signal levels. 
\begin{table}[tb]
	\caption{Evaluation metrics on CHIME-2 test set.}
	\label{tab:results}
	\centering
	\begin{tabular}{l|c|ccccc}
		\toprule
		augmentation & loss	& PESQ & STOI & CD & SI-SDR & fwSegSNR \\\midrule
		- 				& noisy			& 2.29 		& 81.39 	& 5.46 		& 1.92 		& 16.96 \\
		none			& standard		& 3.27		& 91.20		& 2.90		& 9.48		& 23.57 \\
		SNR 			& standard		& 3.31		& 91.40		& \bf{2.85}	& 9.45		& 23.30\\
		SNR+spec	 	& standard		& \bf{3.32}	& 91.57		& 2.89		& 9.55		& 23.30\\
		SNR+spec+level	& standard		& 3.30		& \bf{91.68}& 2.87		& \bf{9.57}	& \bf{23.48}\\
		SNR+spec+level 	& normalized	& 3.31	& 91.55		& 2.89		& 9.52		& 23.41\\\bottomrule
	\end{tabular}
\end{table}
The validation \ac{SI-SDR} shows similar behavior as PESQ.

Tab.~\ref{tab:results} shows the results on the CHIME-2 test set. The enhancement systems improve all results substantially over the noisy input. Adding SNR augmentation adds a gain of 0.05 PESQ. As in the development set, spectral augmentation adds an additional minor improvement. Interestingly, on the test set, the normalized loss shows no influence on the results. We assume this is due to that fact that the given test set does not exhibit largely varying signal levels.

\section{Conclusion}
\label{sec:conclusion}
We have shown the effectivity of data augmentation techniques for supervised deep learning-based speech enhancement by augmenting the SNR, spectral shapes, and signal levels. As level augmentation degrades the performance of the learning algorithm when using level-dependent losses, we proposed a normalization technique for the loss, which is shown to overcome this issue. The experiments were conducted using a real-time capable recurrent neural network on the reverberant CHIME-2 dataset. Future work will also investigate augmentation for acoustic conditions with reverberant impulse responses.

\newpage
%
%
%
\bibliographystyle{splncs04}
\bibliography{sapref.bib}

\begin{thebibliography}{10}
\providecommand{\url}[1]{\texttt{#1}}
\providecommand{\urlprefix}{URL }
\providecommand{\doi}[1]{https://doi.org/#1}

\bibitem{Wang2018}
{Wang}, D., {Chen}, J.: Supervised speech separation based on deep learning: An
  overview. {IEEE/ACM} Trans. Audio, Speech, Lang. Process.  \textbf{26}(10),
  1702--1726 (Oct 2018)

\bibitem{Reddy2020}
Reddy, C.K.A., Beyrami, E., Dubey, H., V., G., Cheng, R., Cutler, R.,
  Matusevych, S., Aichner, R., Aazami, A., Braun, S., P., R., Srinivasan, S.,
  Gehrke, J.: The interspeech 2020 deep noise suppression challenge: Datasets,
  subjective speech quality and testing framework. In: to appear in Proc.
  Interspeech 2020

\bibitem{Ephraim1984}
Ephraim, Y., Malah, D.: Speech enhancement using a minimum-mean square error
  short-time spectral amplitude estimator. {IEEE} Trans. Acoust., Speech,
  Signal Process.  \textbf{32}(6),  1109--1121 (Dec 1984)

\bibitem{Gerkmann2011}
Gerkmann, T., Hendriks, R.C.: Noise power estimation based on the probability
  of speech presence. In: Proc. {IEEE} Workshop on Applications of Signal
  Processing to Audio and Acoustics ({WASPAA}). pp. 145--148 (Oct 2011)

\bibitem{Martin2001}
Martin, R.: Noise power spectral density estimation based on optimal smoothing
  and minimum statistics. {IEEE} Trans. Speech Audio Process.  \textbf{9},
  504--512 (Jul 2001)

\bibitem{Tu2018}
{Tu}, Y.H., {Tashev}, I., {Zarar}, S., {Lee}, C.: A hybrid approach to
  combining conventional and deep learning techniques for single-channel speech
  enhancement and recognition. In: Proc. {IEEE} Intl. Conf. on Acoustics,
  Speech and Signal Processing (ICASSP). pp. 2531--2535 (April 2018)

\bibitem{Valin2018}
{Valin}, J.: A hybrid {DSP}/deep learning approach to real-time full-band
  speech enhancement. In: 20th Intl. Workshop on Multimedia Signal Processing
  (MMSP). pp.~1--5 (Aug 2018)

\bibitem{Tan2018}
Tan, K., Wang, D.: A convolutional recurrent neural network for real-time
  speech enhancement. In: Proc. Interspeech. pp. 3229--3233 (2018)

\bibitem{Wichern2017}
{Wichern}, G., {Lukin}, A.: Low-latency approximation of bidirectional
  recurrent networks for speech denoising. In: Proc. {IEEE} Workshop on
  Applications of Signal Processing to Audio and Acoustics ({WASPAA}). pp.
  66--70 (Oct 2017)

\bibitem{Xia2020}
Xia, R., Braun, S., Reddy, C., Dubey, H., Cutler, R., Tahev, I.: Weighted
  speech distortion losses for neural-network-based real-time speech
  enhancement. In: Proc. {IEEE} Intl. Conf. on Acoustics, Speech and Signal
  Processing (ICASSP) (2020)

\bibitem{Cho2014}
Cho, K., Merri{\"e}nboer, B.V., Bahdanau, D., , Bengio, Y.: On the properties
  of neural machine translation: Encoder-decoder approaches. In: Proc. Eighth
  Workshop on Syntax, Semantics and Structure in Statistical Translation
  (SSST-8) (2014)

\bibitem{Wisdom2019}
{Wisdom}, S., {Hershey}, J.R., {Wilson}, K., {Thorpe}, J., {Chinen}, M.,
  {Patton}, B., {Saurous}, R.A.: Differentiable consistency constraints for
  improved deep speech enhancement. In: Proc. {IEEE} Intl. Conf. on Acoustics,
  Speech and Signal Processing (ICASSP). pp. 900--904 (May 2019)

\bibitem{Loshchilov2019}
Loshchilov, I., Hutter, F.: Decoupled weight decay regularization. In:
  International Conference on Learning Representations (2019),
  \url{https://openreview.net/forum?id=Bkg6RiCqY7}

\bibitem{PESQ_P862}
ITU-T: Recommendation {P.862}: Perceptual evaluation of speech quality
  ({PESQ}), an objective method for end-to-end speech quality assessment of
  narrowband telephone networks and speech codecs (Feb 2001)

\bibitem{Ephrat2018}
Ephrat, A., Mosseri, I., Lang, O., Dekel, T., Wilson, K., Hassidim, A.,
  Freeman, W.T., Rubinstein, M.: Looking to listen at the cocktail party: A
  speaker-independent audio-visual model for speech separation. ACM Trans.
  Graph.  \textbf{37}(4) (Jul 2018)

\bibitem{Wilson2018}
{Wilson}, K., {Chinen}, M., {Thorpe}, J., {Patton}, B., {Hershey}, J.,
  {Saurous}, R.A., {Skoglund}, J., {Lyon}, R.F.: Exploring tradeoffs in models
  for low-latency speech enhancement. In: Proc. Intl. Workshop Acoust. Signal
  Enhancement ({IWAENC}). pp. 366--370 (Sep 2018)

\bibitem{Vincent2013}
Vincent, E., Barker, J., Watanabe, S., Nesta, F.: The second '{CHIME}' speech
  separation and recognition challenge: datadata, tasks and baselines. In:
  Proc. {IEEE} Intl. Conf. on Acoustics, Speech and Signal Processing (ICASSP)
  (June 2012)

\bibitem{Taal2011}
Taal, C.H., Hendriks, R.C., Heusdens, R., Jensen, J.: {An Algorithm for
  Intelligibility Prediction of Time-Frequency Weighted Noisy Speech}. {IEEE}
  Trans. Audio, Speech, Lang. Process.  \textbf{19}(7),  2125--2136 (Sept 2011)

\bibitem{Roux2019}
{Roux}, J.L., {Wisdom}, S., {Erdogan}, H., {Hershey}, J.R.: {SDR} - half-baked
  or well done? In: Proc. {IEEE} Intl. Conf. on Acoustics, Speech and Signal
  Processing (ICASSP). pp. 626--630 (May 2019)

\bibitem{Hu2008}
Hu, K., Divenyi, P., Ellis, D., Jin, Z., Shinn-Cunningham, B.G., Wang, D.:
  Preliminary intelligibility tests of a monaural speech segregation system.
  In: Proc. Workshop on Statistical and Perceptual Audition. Brisbane (Sep
  2008)

\end{thebibliography}

\end{document}